\documentclass[twoside,english,10pt]{JAC2001}

\usepackage{amssymb}
\usepackage{graphicx}
\usepackage{babel}
\usepackage{multicol}
\usepackage[centerlast]{caption2}

\begin{document}

\title{Vibrational Stability of NLC Linac and Final Focus Components}
\author{F.~Le~Pimpec, S.~Adiga (Stanford Univ.), F.~Asiri, G.~Bowden, D.~Dell'Orco,
E.~Doyle, \\ B.~McKee, A.~Seryi ~~SLAC, CA USA; H.~Carter,
C.~Boffo ~~FNAL, IL USA
\thanks{Work supported by the U.S. Department of Energy, Contract
  DE-AC03-76SF00515.}
 }
\maketitle

\begin{abstract}
Vertical vibration of linac components (accelerating structures,
girders and quadrupoles) in the NLC has been studied
experimentally and analytically. Effects such as structural
resonances and vibration caused by cooling water both in
accelerating structures and quadrupoles have been considered.
Experimental data has been compared with analytical predictions
and simulations using ANSYS. A design, incorporating the proper
decoupling of structure vibrations from the linac quadrupoles, is
being pursued.

\end{abstract}


\section{Introduction}
As part of the R\&D effort for the Next Linear Collider (NLC), a
program has developed to study the vibrations induced by cooling
water on the NLC components.

An adequate flow of cooling water to the accelerating structures
is required in order to maintain the structure at the designated
operating temperature. This flow may cause vibration of the
structure and its supporting girder. The acceptable tolerance for
vibration of the structure itself is rather loose $\sim 1 \mu m$.
However our concern is that this vibration can couple to the linac
quadrupoles, where the vibration tolerance is 10~nm, either via
the beam pipe with its bellows or via the supports.

In this paper we will briefly show results obtained for the NLC
RF~structure and girder \cite{lepimpec:Epac02}, and then focus on
vibration of a linac quadrupole, including consideration of
coupling between the structure and the quadrupole.

\section{Vibration of RF~Structure}
\label{RFstructure}

\begin{figure}[tbph]
\begin{center}
\vspace{-0.5cm}
\includegraphics[clip=,totalheight=4.6cm]{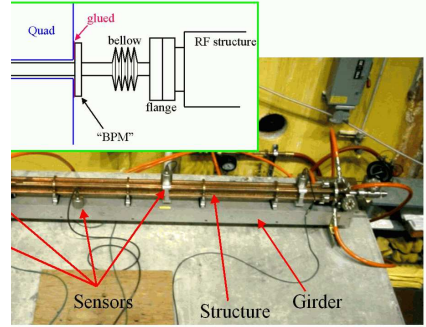}
\end{center}
\vspace{-0.7cm}
\caption{Experimental setup in
the SLD pit: RF structure-Girder system connected to the EM quad.}
\vspace{-0.6cm}
\label{QuadDDSsetup}
\end{figure}

The structure studied is 1.8~m long and is supported by a
``strongback'' (hollow aluminum beam 4x6 inches) of the same
length, Fig.\ref{QuadDDSsetup}. In the design, it was assumed that
3 such structures would be mounted on a single 6~m long girder
\cite{zdr}. The required water flow (at 70MV/m) is about
$\sim$1~$\ell$/s for each structure (in total, through four
cooling copper tubes). The structure was connected to the quad
with a bellow, and a simple mock-up of a BPM was connected (glued)
to the quadrupole. It should be noted that the NLC currently plans
to use shorter RF structures than the one studied
\cite{adolphsen:Epac02}.

\begin{figure}[tbph]
\begin{center}
\vspace{-0.4cm}
\includegraphics[clip=,totalheight=5.3cm]{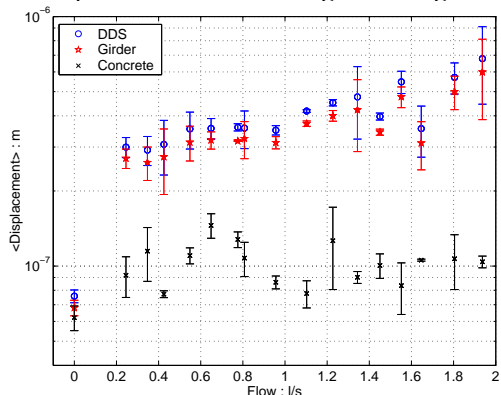}
\end{center}
\vspace{-0.8cm} \caption{Average integrated displacement above 4Hz
of the RF structure (DDS), girder, and of the support (concrete
block) with NLCTA water supply.} \vspace{-0.5cm}
\label{NLCTAflowDDS}
\end{figure}

Fig.\ref{NLCTAflowDDS} displays the results obtained in measuring
the vertical vibration induced by different flow rates passing
through the structure-girder system \cite{lepimpec:Epac02}. Note
that the system considered is above the turbulence threshold
(Re$>$2000) when the flow $>$~0.1~$\ell$/s. In
Fig.\ref{NLCTAflowDDS} the water was supplied by the NLC Test
Accelerator (NLCTA) water system. In this case, the displacement
of the structure-girder is weakly dependant of the flow variation
in the structure because the supplying cooling water has
significant fluctuations of pressure in it (external turbulence).

The NLC cooling system will be designed so that pressure
fluctuations in the cooling water will be limited (if necessary,
by use of passive devices as typically done in industry). Thus,
aiming to understand the contribution to vibration of the internal
turbulence occurring inside the structure itself, we conducted the
second set of experiments. In this case, the structure-girder was
installed in a quieter place on the floor of the SLD (SLAC Large
Detector) collider hall and the cooling water was gravity-fed from
a tank located $\sim$18~m higher. The structure-girder was bolted
to a $\sim$26T concrete block initially placed on a rubber mat and
then on sand (in the first configuration the block had resonance
at $\sim$35Hz which was decreased in the second case). The
vibrations were monitored either by piezo-accelerometers or by
seismometers and one piezo-transducer was used to measure water
pressure fluctuations. In both sets of experiments (NLCTA and SLD)
the flows in four cooling tubes were in opposite direction (2 by
2).

\begin{figure}[tbph]
\begin{center}
\vspace{-0.3cm}
\includegraphics[clip=,width=5.3cm]{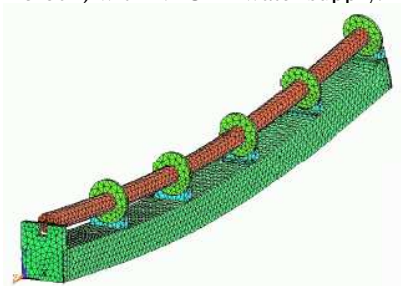}
\end{center}
\vspace{-0.9cm} \caption{ANSYS simulation of the RF structure and
Al girder, showing the lowest resonance mode.}
\vspace{-0.6cm}
\label{ANSYS49hz}
\end{figure}

We have shown in \cite{lepimpec:Epac02} that the vibration
spectrum of the girder-structure system exhibits a vertical
resonance at $\sim$52Hz. Simulations using ANSYS code have shown
that the natural first resonant frequency for such design is about
$\sim$49~Hz, in good agreement with measurements, and corresponds
to simplest vertical bending mode Fig.\ref{ANSYS49hz}. These
simulations also indicate that the second and the third modes are
the horizontal dipole at $\sim$69~Hz and vertical two-nodes mode
at $\sim$117~Hz, while the fourth resonance is torsional
$\sim$146~Hz. The driving forces (ground motion, pressure DP/P, …)
decrease rapidly with frequency. One possibility to further reduce
the vibration of the structure-girder system is to design a girder
which has a higher first resonant frequency. For further studies,
we have set a goal of increasing the lowest resonance frequency to
130~Hz and performed simulations to understand what modifications
this would require. One way to stiffen the girder is to increase
its dimensions. Simulations have shown that keeping the same
material and design but increasing the girder size (6"x4" to
10"x10") and the wall thickness (from 0.25" to 1") lead to a
lowest natural frequency of 120~Hz. Such big increase of the
resonance frequency may not be necessary, but the studies have
shown that significant improvement is possible with simple
modification of the girder design.

\section{Vibration of RF structure and Coupling to Quadrupole }

Using the setup of Fig.\ref{QuadDDSsetup} we have studied the
vibration of RF structure versus flow, and the coupling of
vibration from the RF structure to the EM quadrupole in the case
when RF structure is cooled with gravity-fed water.

\begin{figure}[tbph]
\begin{center}
\vspace{-0.3cm}
\includegraphics[clip=,totalheight=5.3cm]{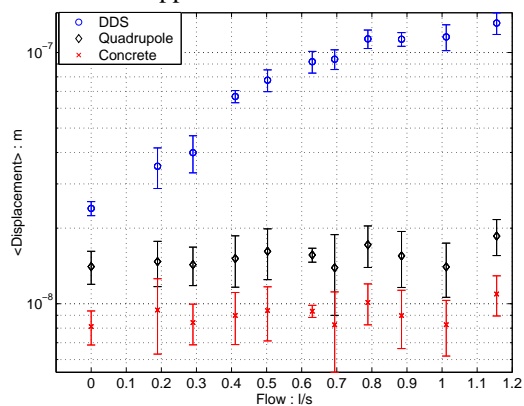}
\end{center}
\vspace{-0.7cm}
\caption{Average integrated displacement above 4Hz,
with vacuum and gravity fed water.} \vspace{-0.4cm}
\label{SLDflowVac}
\end{figure}

\begin{figure}[tbph]
\begin{center}
\vspace{-0.3cm}
\includegraphics[clip=,totalheight=5.3cm]{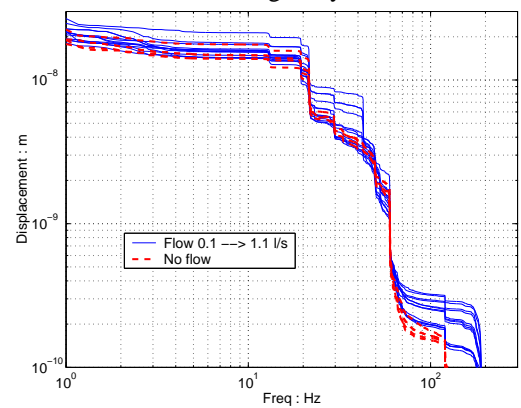}
\end{center}
\vspace{-0.6cm} \caption{Quadrupole integrated displacement with
four different flows in the RF-structure, -SLD measurement.}
\vspace{-0.2cm} \label{quadSLDspectrum}
\end{figure}

\begin{figure}[tbph]
\begin{center}
\includegraphics[clip=,totalheight=5.3cm]{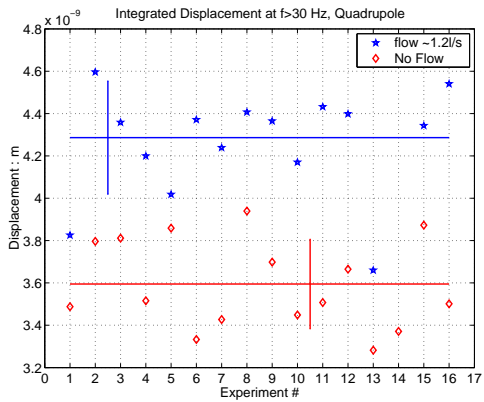}
\end{center}
\vspace{-0.7cm}
\caption{Coupling from the water cooled
RF-structure to the Quadrupole above 30Hz.}
\vspace{-0.6cm}
\label{quadcoupling30hz}
\end{figure}

Vibration of the RF structure versus flow is shown in
Fig.\ref{SLDflowVac}. In this case,  vibrations are caused mostly
by the internal turbulence occurring in the RF structure. At
nominal flow $1 \ell/s$ vibration of the structure is $\sim$110nm,
in comparison with 350nm obtained with NLCTA cooling water
\cite{lepimpec:Epac02}. Additional vibrations of the quadrupole
are small. Performing multiple measurements with and without flow,
and analyzing spectra of quadrupole vibration
(Fig.\ref{quadSLDspectrum}), we found that the additional
vibration of the quadrupole due to cooling of RF structure above
30Hz is 2.4nm (obtained as $(4.3^2-3.6^2)^{0.5}$, assuming
vibrations are uncorrelated), see Fig.\ref{quadcoupling30hz}.
Taking lower cut frequency would be statistically uncertain, due
to high background noise. These results suggest that coupling from
RF structure to the quadrupole is about 2\% in the current
configuration. We also investigated influence of vacuum in the RF
structure (and possible stiffening of the bellow) on this
coupling. No noticeable difference was observed with or without
vacuum (the results displayed in Fig.\ref{SLDflowVac} are obtained
with a primary vacuum of about ~10$^{-1}$~Torr in the
structure-quadrupole system). However, we have not yet studied how
much coupling is due to the bellow connection and how much due to
transmission via support and concrete.

One should also note that the present set up is simplified. In
particular, the quadrupole was placed on small granite stand (with
shims to adjust the height), which was placed on concrete block
(without rigid connections). Such system had amplification -- the
quadrupole vibration is higher than the concrete as seen in
Fig.\ref{SLDflowVac}. This can be avoided in real system.

\section{Vibration of EM Quadrupole}
\label{Quadresults}

\begin{figure}[htbp]
\begin{center}
\vspace{-0.2cm}
\includegraphics[clip=,totalheight=6cm, width=4.5cm]{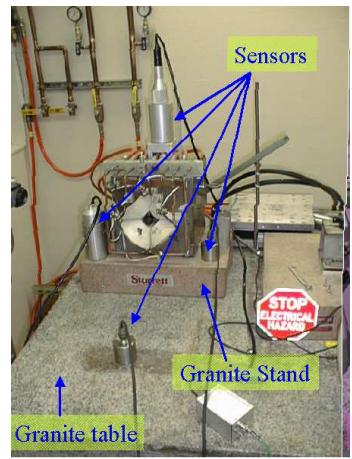}
\end{center}
\vspace{-0.7cm}
\caption{EM quadrupole vibration measurement setup. }
\vspace{-0.2cm}
\label{Quadsetup}
\end{figure}

The NLC project calls for maximal use of permanent magnet (PM)
quadrupoles which will not have cooling water. The electromagnet
quadrupoles (EM) however are also prototyped for NLC and we
studied vibration caused by cooling water in such EM quadrupole.
The EM quadrupoles was fed by a standard water supply for a
nominal flow of $\sim$0.1~$\ell/s$ obtained with pressure
difference of 8.5 bar. The quadrupole was installed on a granite
table Fig.\ref{Quadsetup}. The table was installed on rubber pads
to isolate the table from the high frequency vibration in the
noisy environment where measurements were performed. This reduced
the high frequency background, but significantly amplified
frequencies below 6-9~Hz, making it possible to study the effect
of cooling water on quadrupole vibration only above about 10Hz.

For f$>$20Hz, the vibration induced by the flow of
$\sim$0.1~$\ell/s$ in the quad is roughly 3.5~nm$\pm $0.25~nm
while 1~nm$\pm $0.25~nm at rest (averaged on several
measurements). Assuming that the additional vibration is
uncorrelated, the effect due to cooling water itself is:
$\sqrt{(3.52^2 - 1)}$ = 3.35~nm. The result is similar if a lower
cut frequency (e.g. 15Hz) was considered, until below 10Hz where
statistical error becomes too big. Note that earlier studies of
FFTB quadrupole stability \cite{fftb_quad_vibro} have shown that
the effect of the cooling water is on a nanometer level as well,
for quadrupoles that were (in contrast to our study) also properly
placed on movers.

With these data, we can estimate that in the pessimistic case, if
the cooling water will be similar to NLCTA (with similar pressure
fluctuations), vibration of the quadrupoles will scale to about
7.6nm due to coupling to the RF structures. In the case of EM
quadrupoles, there will be about 3.3nm additional due to cooling
of the quadrupoles themselves, which in total amounts to
$\sqrt{(7.6^2 +3.3^2)}$ = 8.3nm. This value is below the tolerance
but has little margin. However, simple design optimizations,
discussed above, are expected to reduce these numbers
considerably.

\begin{figure}[tbph]
\begin{center}
\vspace{-0.2cm}
\includegraphics[clip=,totalheight=5.3cm]{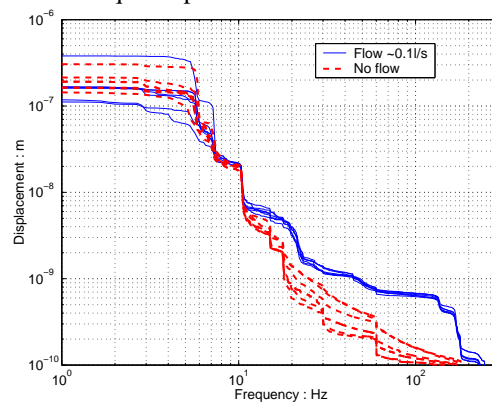}
\end{center}
\vspace{-0.7cm}
\caption{Integrated Displacement Spectrum of the
water induced vibration in the EM quadrupole at nominal flow. The
region below 10Hz is disturbed by resonances of the concrete table
installed on rubber pads. }
\vspace{-0.5cm}
\label{Quadresltspectrum}
\end{figure}

Among further studies of RF structure and quadrupole vibration
planned at FNAL and SLAC are: performing measurements in quieter
place, to quantify lower frequency range; study the case of
quadrupole been placed on movers and realistic independent
supports; continuing optimization of the system as a whole.

\section{Conclusion}

Cooling water can cause vibration of an accelerating structure
both through internal turbulence in the cooling pipes on the
structure, and through pressure fluctuations in the supply water
(external turbulence) \cite{lepimpec:Epac02}. The latter does not
depend on the flow rate through the structure and can be the
dominant source of vibration in practical situations. For the case
studied, mechanical resonances of the structure-girder assembly
explain the measured amplitudes. Optimization of design to
increase resonance frequencies is expected to reduce vibration.
Coupling from RF structure to linac quadrupoles can occur via
bellows and the support but was measured to be at the percent
level. Present studies suggest that the vibration tolerances for
the NLC linac quadrupoles are met, but without much margin.
Optimization of the girder design to improve its vibration
property is highly desirable and will be pursued.

\section{Acknowledgments}
We would like to thank R. Assmann, M. Breidenbach,  T.
Raubenheimer, S. Redaelli, N. Solyak and C. Spencer for help and
useful discussions.

\end{document}